\documentclass[10pt,conference]{IEEEtran}
\usepackage{booktabs,cite,comment,flushend,listings,multirow,tikz,todonotes}
\usepackage{graphicx,caption,subcaption}
\usetikzlibrary{arrows,calc,shapes}
\usepackage{amsmath} 
\usepackage{amssymb}
\usepackage{numprint}

\usepackage[utf8]{inputenc}
\usepackage[T1]{fontenc}
\usepackage{microtype}

\usepackage{tabularx}
\usepackage{hyperref}

\DeclareGraphicsExtensions{.pdf,.png}
\graphicspath{{img/}}

\newcommand{\dissertation}[1]{\iffalse{#1}\fi}
\newcommand{\ed}[1]{\textcolor{red}{#1}}
\newcommand{\seedtwo}[1]{\textcolor{black}{#1}}
\newcommand{\seedthr}[1]{\textcolor{black}{#1}}

\newcommand{\matias}[1]{\todo[color=green!75,inline]{#1}}

\newcommand{\td}{\textcolor{red}{TODO}~}

\makeatletter
\newcommand*{\defeqq}{\mathrel{\rlap{%
                     \raisebox{0.3ex}{$\m@th\cdot$$\m@th\cdot$}}%
                     \raisebox{-0.3ex}{$\m@th\cdot$$\m@th\cdot$}}%
                     =}
\makeatother

\begin{document}
\title{Sorting~and~Transforming~Program~Repair\\Ingredients~via~Deep~Learning~Code~Similarities}

\author{
\IEEEauthorblockN{Martin~White\IEEEauthorrefmark{1},~Michele~Tufano\IEEEauthorrefmark{1},~Mat{\'i}as~Mart{\'i}nez\IEEEauthorrefmark{2},~Martin~Monperrus\IEEEauthorrefmark{3},~and~Denys~Poshyvanyk\IEEEauthorrefmark{1}}
\IEEEauthorblockA{\IEEEauthorrefmark{1}College~of~William~and~Mary,~Williamsburg,~Virginia,~USA\\Email:~\{mgwhite,~mtufano,~denys\}@cs.wm.edu}
\IEEEauthorblockA{\IEEEauthorrefmark{2}Université~Polytechnique~Hauts-de-France,~Valenciennes,~France\\Email:~matias.martinez@uphf.fr}
\IEEEauthorblockA{\IEEEauthorrefmark{3}KTH~Royal~Institute~of~Technology,~Stockholm,~Sweden\\Email:~martin.monperrus@csc.kth.se}
}
\maketitle
\begin{abstract}
In the field of automated program repair, the {redundancy assumption} 
claims large programs contain the seeds of their own repair.~%
However, most redundancy-based program repair techniques do not reason about the repair ingredients---the code that is reused to craft a patch.~%
We aim to reason about 
the repair ingredients by using code similarities to prioritize \emph{and transform} statements in a codebase for patch generation.~%
Our approach, DeepRepair, relies on deep learning to reason about code similarities.~
Code fragments at well-defined levels of granularity in a codebase can be sorted according to their similarity to suspicious elements (i.e., code elements that contain suspicious statements) and \emph{statements can be transformed} by mapping out-of-scope identifiers to similar identifiers in scope.~%
We examined these new search strategies for patch generation with respect to effectiveness from the viewpoint of a software maintainer.~%
Our comparative experiments were executed on six open-source Java projects including 374 buggy program revisions and consisted of {19,949} trials spanning {2,616} days of computation time.~%
DeepRepair's search strategy using code similarities generally found compilable ingredients faster than the baseline, jGenProg, but this {improvement} neither yielded test-adequate patches in fewer attempts (on average) nor found significantly more patches (on average) than the baseline.~%
Although the patch counts were not \emph{statistically} different, there were notable differences between the nature of DeepRepair patches and 
jGenProg patches.~%
The results 
show that our learning-based approach finds patches that \emph{cannot} be found by existing redundancy-based repair techniques.~
\end{abstract}
\section{Introduction}
\label{sec:introduction}

In the field of automated program repair, many repair approaches are based on a common intuition:~a 
patch can be composed of source code residing elsewhere in the repository or even in other projects.~%
These approaches are called redundancy-based repair techniques, since they leverage redundancy and repetition in source code~\cite{Pierret:2009,Gabel:2010,Hindle:2012,Carzaniga:2013,Nguyen:2013:ASE,Martinez:2014,Barr:2014}.~%
For example, GenProg~\cite{LeGoues:2012:TSE,LeGoues:2012} reuses existing code from the same codebase, whereas CodePhage~\cite{Sidiroglou-Douskos:2015} transplants checks from one application to another.~
This intuition has been examined by at least two independent empirical studies, showing that a significant proportion of commits are indeed composed of existing code~\cite{Martinez:2014,Barr:2014}.~

The code that is reused to craft a patch is called a \emph{repair ingredient}.~%
Most redundancy-based repair techniques harvest repair ingredients at random and do not reason further about the optimal strategies to select repair ingredients.~%
In other words, these repair techniques simply brute-force {search} for {viable} ingredients, randomly searching the codebase in a straightforward trial-and-error process.~%
Although these techniques are able to find 
patches, 
their naive search and rigid application of repair ingredients means \emph{patches that use novel expressions are unattainable}.~%

In this paper, we aim to improve the reasoning about repair ingredients before they enter the repair pipeline (i.e., ingredient insertion, compilation, and test execution).~%
Our key intuition is that a good ingredient does not come from just \emph{any} location in the program but comes from \emph{similar} code.~%
We design our experiments to evaluate whether an approach built on this intuition can effectively improve patch generation.~%

We rely on deep learning to reason about code similarities.
Deep learning has provided the software engineering research community with new ways to mine and analyze data to support tasks~\cite{White:2015,Corley:2015,Yang:2015,Lam:2015,White:2016,Wang:2016,Gu:2016,Tufano:2018,Tufano:2018:ASE,Tufano:2019}.~%
Our learning-based approach automatically creates a representation of source code that accounts for the structure \emph{and meaning} of lexical elements.~%
Our approach is completely unsupervised, and no top-down specification of features is made beforehand.~%
We use the learned features to compute distances between code elements to measure similarities, and we use the similarities to intelligently select \emph{and adapt} repair ingredients 
to generate a series of statement-level edits for program repair.~%

We design, implement, and evaluate DeepRepair, an approach for sorting and transforming program repair ingredients via deep learning code similarities.~%
Our approach is implemented on top of Astor~\cite{Martinez:2016:ASTOR}, a Java implementation of GenProg.~%
DeepRepair uses recursive deep learning~\cite{Socher:2014} to prioritize repair ingredients in a fix space and to select code that is most similar when choosing a repair ingredient 
for patch generation.~%
\emph{Additionally, DeepRepair addresses a major limitation of current redundancy-based repair techniques that do not {adapt} repair ingredients by automatically transforming ingredients based on {lexical elements'} similarities}.~%
For instance, given the simple repair ingredient
\texttt{\textcolor{blue}{return} FastMath.abs(y-x) <= eps;}
the transformation may involve replacing the variable eps (which is out of scope at the {modification point}) with the variable SAFE\_MIN (
in scope at the modification point).~

We assess the effectiveness of using code similarities to sort or transform repair ingredients 
for patch generation.~%
To do so, we evaluate our approach along several dimensions to measure what aspects contribute and do not contribute to improved repair effectiveness.~%
We compute a baseline; evaluate sorting (at different levels of granularity and scope) in isolation; evaluate transforming in isolation (at different levels of granularity and scope); and evaluate sorting with the ability to transform {repair ingredients} (at different levels of granularity and scope).~%
We evaluate everything on six open-source Java projects including 374 buggy program revisions in Defects4J database version {1.1.0}~\cite{Just:2014,Just:2014:ISSTA}.~%
In summary, our evaluation consists of {19,949} trials spanning {2,616} days of computation time.~%
DeepRepair's search strategy using code similarities generally found compilable ingredients (Sec.~\ref{sub:repair}) faster than the baseline, jGenProg~\cite{Martinez:2016}, but this improvement neither yielded test-adequate patches in fewer attempts (on average) nor found 
significantly more patches (on average) than the baseline.~%
Although the patch counts were not \emph{statistically} different, there were notable differences between the nature of DeepRepair's patches and jGenProg's patches.~
To sum up, we make the following noteworthy contributions:~%
\begin{itemize}
\item
a novel 
learning-based algorithm to intelligently select and {adapt} repair ingredients for redundancy-based repair;~%
\item
the implementation of our approach for Java in a publicly available tool;~
\item
an evaluation protocol for redundancy-based repair, including a set of novel metrics that are specific to the analysis of ingredient selection strategies;~%
\item
an evaluation on 374 real bugs from the Defects4J bench-mark showing that our algorithm finds patches that cannot be found by
existing redundancy-based techniques;~%
\item
online appendix with code and experimental results~\cite{onlineappendix}.
\end{itemize}
The paper is organized as follows.~%
Sec.~\ref{sec:related} reviews background on program repair techniques and related work on the redundancy assumption. 
Our approach is based on two empirically validated results:~redundancy is localized in the same file and many repairs do not use new tokens.~%
Sec.~\ref{sec:approach} presents DeepRepair, 
the first approach that can sort the fix space at well-defined levels of granularity (e.g., methods and classes) and transform program repair ingredients.~%
Sec.~\ref{sec:validation} specifies our experimental design.~%
Sec.~\ref{sec:results} reports our results.~%
Like typical repair studies, we report the number of bugs unlocked by our approach.~%
In our study, unlocking bugs is attributed to DeepRepair's ability to expand the fix space by transforming ingredients.~%
On the other hand, to the best of our knowledge, unlike any repair study conducted to date, we also report the statistical significance of the difference in patch counts/attempts between our approach and the baseline.~%
In summary, DeepRepair not only patches bugs that cannot be patched by existing generate-and-validate techniques but also finds sets of test-adequate patches that are distinctly different than the baseline's patches.~%
The intent of our statistical analysis was to responsibly qualify the key result that DeepRepair unlocks new bugs.~%
Sec.~\ref{sec:threats} discusses threats to the validity of our work.~%
Sec.~\ref{sec:conclusion} concludes the paper.~%
\section{Background~and~Related~Work}
\label{sec:related}

\subsection{Automated~Program~Repair}
\label{sub:automated}

Automated 
repair involves the transformation of an unacceptable behavior of a program execution into an acceptable one according to a specification~\cite{Monperrus:2015}.~%
Behavioral repair techniques 
change the behavior of a program under repair by changing its 
source or binary code~\cite{Monperrus:2015}.~%
For example, GenProg~\cite{LeGoues:2012:TSE,LeGoues:2012,LeGoues:2012:GECCO}, which was used in several studies~\cite{DBLP:journals/tse/GouesHSBDFW15, DBLP:conf/sigsoft/YiAKTR17}, changes the behavior of a program 
according to a {test suite} (i.e., an input-output specification) by modifying the program's source code.
This \emph{generate-and-validate} technique searches for statement-level modifications to make to an abstract syntax tree (AST).~%
%
%

A complementary set of repair techniques leverage program analysis and program synthesis {to repair programs by constructing code with particular properties}~\cite{Nguyen:2013:ICSE,Mechtaev:2015,Xuan:2016,Mechtaev:2016, Tian:2017:ADR:3106237.3106300, Le:2017:SSS:3106237.3106309}.~%
For example, SemFix~\cite{Nguyen:2013:ICSE} {synthesizes} (side-effect free) expressions for replacing the right-hand side of assignments or branch predicates to repair programs.~%
Angelix~\cite{Mechtaev:2016} uses guided symbolic execution and satisfiability modulo theories solvers to {synthesize} patches using \emph{angelic} values, i.e., expression values that make a given test case pass.~%
Nopol~\cite{Xuan:2016} either modifies an existing conditional expression or adds a precondition to a statement or block in the code.~%

Both types of techniques have distinct advantages.~%
Generate-and-validate techniques have the advantage of operating at coarse granularity with the power to mutate statements.~%
{However, these techniques generally do not mutate code below statement-level granularity, so they do not change conditional expressions nor variables}.~%
Semantics-based techniques have the advantage of operating at fine granularity on expressions and variables, enabling them to synthesize a repair even if the patch code does not exist in the codebase~\cite{Nguyen:2013:ICSE}.~%
However, they generally do not operate at higher levels of granularity, and scalability has been a key concern.~%

{SearchRepair~\cite{Ke:2015} draws from both generate-and-validate and semantics-based techniques.}~%
{SearchRepair uses symbolic reasoning to search for code and semantic reasoning to generate candidate patches.}~%
{Consequently, scalability is a concern, and SearchRepair has only been 
shown to work on small programs.}~%
{In lieu of program semantics to search for similar code, we use a learning-based approach to query textually/functionally similar code at arbitrary levels of granularity.
SearchRepair uses several software systems to repair small, student programs, whereas our approach 
uses the program under repair, and we aim to repair real software systems.~%
Finally, SearchRepair depends on input-output examples to describe desired behavior, whereas we automatically learn features for distinguishing code fragments.~%

Recently, Yokoyama et al.~\cite{Yokoyama:2016} used code similarity to select code lines in code regions similar to the faulty code regions.~%
Our work using code similarities differs in several important respects.~%
They used small, fixed-sized code regions of 4, 6, and 8 lines, whereas we use code regions at arbitrary levels of granularity, making it possible to map a suspicious statement (i.e., a statement suspected to contain a bug~\cite{Martinez:2014:ASTOR}) to its method or class and query similar execution contexts or similar classes.~%
Their similarity metric analyzed the longest common subsequence between two token sequences, whereas we use a learning-based clone detector that fuses information on structure and identifiers and is capable of finding more meaningful similarities than token-based techniques~\cite{White:2016}.~%
They used a collection of 24 bug-fix commits and defined a code coverage metric to evaluate their approach.~%
We use a collection of {374} reproducible bugs and implemented our learning-based approach in an automatic software repair framework to measure effectiveness.~%
Lastly, our approach is capable of using similarities to \emph{transform repair ingredients}.~

Our approach relies 
on machine learning for the {fix localization problem}.~%
Prophet~\cite{Long:2016} is a learning-based approach that uses explicitly designed code features to rank candidate repairs.
We use representation learning~\cite{Bengio:2012,LeCun:2015} to automatically learn how to encode fragments to detect similarities.~%
Other approaches train on correct (student) solutions to specific programming tasks and try to learn task-specific repair strategies~\cite{Bhatia:2016,Pu:2016}.~%
For example, in massively open online courses, the programs are generally small and synthetic~\cite{Gupta:2017}.~%
Other approaches cannot transform statements.~%
For example, Gupta et al.~\cite{Gupta:2017} use an oracle that rejects a fix if it does not preserve the identifiers and keywords present in the original statement.~%
We use similarities to map the set of out-of-scope variables to a set of variables in scope at the modification point.~%

\subsection{Redundancy~Assumption}
\label{sub:redundancy}

Martinez et al.~\cite{Martinez:2014} 
examined a critical assumption of GenProg that certain bugs can be fixed by copying and rearranging existing code.~%
They validated the redundancy assumption by defining a concept of \emph{software temporal redundancy}.~%
A commit is temporally redundant if it is only a rearrangement of code in previous commits.~%
They measured redundancy at two levels of granularity: line- and token-level.~
At line-level granularity, they found 
most of the temporal redundancy to be localized in the same \emph{file}.~%
We use 
learning-based code clone detection, which is
capable of detecting \emph{file}-level clones, so the search will first look in the same file and \emph{similar~files}.~%
Moreover, their token-level granularity results imply that many repairs need never invent a new token.~%
Ergo, the tokens exist, but repair engines need to \emph{learn} how to use them.~%
Again, our key insight is to look to the \emph{learning}-based code clone detection approach, which maps tokens to continuous-valued vectors called \emph{embeddings} that we can use to measure similarities.~%
Then we use the similarities to consider different tokens in different contexts.~%
Finally, we draw one more bit of insight from their empirical study.~%
The authors note a tension between working with the line pool or the token pool~\cite{Martinez:2014}, and they characterize this tension by the combination spaces of line- (smaller combination space) versus token-level (larger combination space) granularity.~%
The essence of our work is to suppose there exists a manifold 
governed by coarse-grained snippets like lines yet can be parameterized by fine-grained snippets like tokens.~%
DeepRepair's ingredient transformation traverses this manifold to find repairs that \emph{cannot} be found by existing redundancy-based techniques.~%

Barr et al.~\cite{Barr:2014} examined a history of 15,723 commits to determine the extent to which the commits can be reconstructed from existing code.~%
The grafts they found were mostly single lines, i.e., micro-clones, and they proposed that micro-clones are useful since they are the atoms of code construction~\cite{Barr:2014}.~%
The learning-based clone detection approach uses these micro-clones to compute representations for fragments at higher levels of granularity, which we 
use to assess similarities and 
prioritize statements and values to assign to code modifications.~

\section{Technical~Approach}
\label{sec:approach}

Our approach is organized as a pipeline comprising three phases: recognition, learning, and repair.~%
The language recognition phase (Sec.~\ref{sub:recognition}) consumes source code of the application under repair in situ and produces training data.~%
The machine learning phase (Sec.~\ref{sub:learning}) consumes the training data and produces encoders for encoding anything from a lexical element to a class such that similarities can be detected among the encodings.~%
The program repair phase (Sec.~\ref{sub:repair}) uses the encoders to query and transform code fragments in the codebase for patch generation.~%
To explain our approach, we use bug Math-63 (Revision ID:~d2a5bc0) from the Defects4J database version {1.1.0}~\cite{Just:2014,Just:2014:ISSTA} as a running example throughout 
Sec.~\ref{sec:approach}.~%

\subsection{Language~Recognition~Phase}
\label{sub:recognition}
Our approach begins with a program and its set of source code
files.
%
The first stage of our language recognition phase consumes this source directory and parses its contents to create an AST or \emph{any} model for representing the code.~%
A well-formed, typed AST is sufficient since the visitor design pattern facilitates 
queries on the program under analysis such as the number of files, classes, and methods in the application sources.~%
For example, for the Math~\cite{commons-math} library, 
this stage produces a model containing {459} files, {661} classes, and {4,983} methods.~%

The second stage of our language recognition phase consumes the model and uses a program processor to produce corpora at different levels of granularity.~%
In this context, a program processor is simply a utility for performing a specific action.~%
First, we create a file-level corpus by querying the model for all the files.~%
Our program processor creates one line per code file in the corpus by printing (from left to right) the terminal symbols~\cite{Aho:2006} of the corresponding syntax tree.~%
Simultaneously, we store keys that uniquely identify the Java source code files.~%
Likewise, we use the program processor to build corpora at other levels of granularity, e.g., classes or methods, provided there is a way to uniquely identify the code elements.~%
For example, Math files can be identified by their path; classes can be identified by their qualified name; and methods can be identified by concatenating the qualified name of their parent type and their signature.~%
We build corpora at different levels of granularity because (in the next phase of our approach) we mine patterns at the level of classes, methods, and even lexical elements.~%
These representations will be the mechanisms for querying and transforming ingredients.~%

The third stage of our language recognition phase normalizes the corpora by mapping some or all of the literal tokens to their respective type.~%
For example, every floating point number in the Math corpus would be replaced by a generic symbol for floating point literal values.~%
Normalizing the corpora is the last stage of recognition and staging data for learning.~%

\subsection{Machine~Learning~Phase}
\label{sub:learning}
The first stage of our machine learning phase involves training a neural network language model from the file-level corpus.
A (statistical) language model is a probability distribution over sentences (e.g., lines of code) in a language~\cite{Jurafsky:2009}.~%
Recently, language modeling has been used for software engineering tasks~\cite{Hindle:2012,Afshan:2013,Nguyen:2013:FSE:2,Nguyen:2014,Tonella:2014,Campbell:2014,Tu:2014,Nguyen:2014:ASE,Raychev:2014,Franks:2015,White:2015,Hellendoorn:2015,Ray:2016,Nguyen:2016,White:2016,Wang:2016:ASE,Hellendoorn:2017:DNN:3106237.3106290}.~%
We require a neural network to learn representations for each term in the file-level corpus~\cite{Bengio:2003}, ~
and recurrent neural networks in particular can serve as effective architectures for language models of both natural language corpora and source code~\cite{Mikolov:2010,Mikolov:2011:RNNLM, Mikolov:2011:Extensions,Mikolov:2011:Strategies,Mikolov:2012,White:2015,White:2016}.~%
These models learn from the order of terms in a corpus, imputing 
\emph{embeddings} to the terms in such a way that terms used in similar ways have embeddings that are close to each other in a feature space.~%
We use the embeddings to initialize the second learning stage.~

\begin{figure}[t]
\centering
\input{lst/method.tex}
\vspace{-0.5\baselineskip}
\caption{Suspicious method in Math-63's MathUtils.java}
\vspace{-1.5\baselineskip}
\label{fig:method}
\end{figure}

The second stage of our machine learning phase involves training another learner to encode arbitrary streams of embeddings.~
We leverage work on recursive autoencoders~\cite{Socher:2011:EMNLP,Socher:2013} and learning-based code clone detection~\cite{White:2016}.~%
To demonstrate the 
recursive learning procedure, consider a suspicious method in Math-63's class MathUtils (Fig.~\ref{fig:method}).~%
The method's return statement is filtered to the stream of terms in Fig.~\ref{fig:ast.encode.greedy}.~%
Then our first machine learning stage maps the stream of terms to a stream of embeddings 
$\{x_0,\dots,x_7\}$ (Fig.~\ref{fig:ast.encode.greedy}).~%
There are seven pairs of adjacent terms in Fig.~\ref{fig:ast.encode.greedy}.~%
Each pair of adjacent terms are 
encoded by agglutinating the embeddings $x=[x_\ell;x_r]\in\mathbb{R}^{2n}$ multiplying $x$ by a matrix $\varepsilon=[\varepsilon_\ell,\varepsilon_r]\in\mathbb{R}^{n\times 2n}$ adding a 
bias vector $\beta_z\in\mathbb{R}^n$ 
and passing the result to a nonlinear vector 
function~$f$:~%
\begin{equation}
\label{eq:z}
z=f\left(\varepsilon x+\beta_z\right)
\end{equation}
For example, in Fig.~\ref{fig:ast.encode.greedy}, $x_\ell$ and $x_r$ may correspond to $x_5$ and $x_6$, respectively.~%
The result $z$ represents an encoding for the stream of two terms corresponding to $x$, e.g., ``y x'' in Fig.~\ref{fig:ast.encode.greedy}.~
Then $z$ is 
decoded by multiplying it by a matrix $\delta=[\delta_\ell;\delta_r]\in\mathbb{R}^{2n\times n}$ and adding a 
bias vector $\beta_y\in\mathbb{R}^{2n}$, i.e.,~
\begin{equation}
\label{eq:y}
y=\delta z+\beta_y
\end{equation}
The output $y=[\hat{x}_\ell;\hat{x}_r]\in\mathbb{R}^{2n}$ is referred to as the model's \emph{reconstruction} of the input.~%
This model $\theta=\{\varepsilon,\delta,\beta_z,\beta_y\}$ is called an \emph{autoencoder}, and training the model involves measuring the 
error between the original input vector $x$ and the {reconstruction} $y$, i.e.,~%
\begin{equation}
\label{eq:error}
E(x;\theta)=||x_\ell-\hat{x}_\ell||_2^2+||x_r-\hat{x}_r||_2^2
\end{equation}
Concretely, the model is trained by minimizing Eq.~\eqref{eq:error}.~%
Training the model to encode streams with more than two terms requires recursively applying the autoencoder.~%
To this end, we use the greedy procedure defined by Socher et al.~\cite{Socher:2011}.~%
In the first iteration of the procedure, each pair of adjacent terms are encoded (Fig.~\ref{fig:ast.encode.greedy}).~%
The pair whose encoding yields the lowest reconstruction error (Eq.~\eqref{eq:error}) is the pair selected for encoding at the current iteration.~%
For example, in Fig.~\ref{fig:ast.encode.greedy}, the pair ``y x'' is selected to be encoded first.~%
As a result, in the next iteration, $x_5$ and $x_6$ are replaced by $z$ and the procedure repeats.~%
Upon deriving an encoding for the entire stream, the backpropagation through structure algorithm~\cite{Goller:1996} computes partial derivatives of the (global) error function w.r.t.~$\theta$.~%
Then the error signal is optimized using standard methods.~%

Our intent behind using this learning-based approach is manyfold.~%
First, given that we aimed to assess the effectiveness of using similarities in sorting and transforming fix space elements for patch generation, this recursive learning procedure gave us the ability to evaluate similarity-based sorts at well-defined levels of granularity.~%
The same model can be used to recognize similarities among classes, methods, or even identifiers.~%
Second, the approach did not require intuition in the matter of engineering features for fragments since the approach automatically searched for empirically-based features.~%
Third (and perhaps most important), the learning procedure has been shown to go beyond syntactic similarity~\cite{White:2016}.~%
Aside from syntax implicitly organizing the lexemes, the procedure is not syntax directed.~%
In fact, the greedy procedure is free to encode streams of lexemes that are \emph{not} valid programming constructs.~%
Moreover, by initializing the procedure with trained embeddings, it is capable of recognizing similarities among fragments that are only weakly syntactically similar or even syntactically dissimilar fragments with similar functionality (i.e., Type IV clones)~\cite{White:2016}, which has important implications on prioritizing both textually \emph{and functionally} similar fragments when sorting ingredients.~%

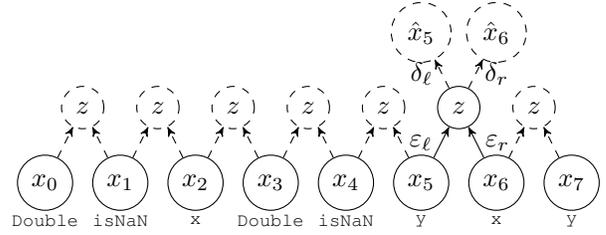
\begin{figure}[t]
\centering
\begin{tikzpicture}[->,>=stealth']
\tikzstyle{neuron}=[circle,draw=black]
\tikzstyle{input}=[neuron,fill=white];
\tikzstyle{hidden}=[input];
\tikzstyle{output}=[neuron,fill=white,dashed];
\node[input] (x-1) at (0,0) {$x_0$};
\node[input] (x-2) at (1,0) {$x_1$};
\node[input] (x-3) at (2,0) {$x_2$};
\node[input] (x-4) at (3,0) {$x_3$};
\node[input] (x-5) at (4,0) {$x_4$};
\node[input] (x-6) at (5,0) {$x_5$};
\node[input] (x-7) at (6,0) {$x_6$};
\node[input] (x-8) at (7,0) {$x_7$};
\node[output] (z-1) at ($(0,1)!0.5!(1,1)$) {$z$};
\node[output] (z-2) at ($(1,1)!0.5!(2,1)$) {$z$};
\node[output] (z-3) at ($(2,1)!0.5!(3,1)$) {$z$};
\node[output] (z-4) at ($(3,1)!0.5!(4,1)$) {$z$};
\node[output] (z-5) at ($(4,1)!0.5!(5,1)$) {$z$};
\node[hidden] (z-6) at ($(5,1)!0.5!(6,1)$) {$z$};
\node[output] (z-7) at ($(6,1)!0.5!(7,1)$) {$z$};
\draw[dashed,->] (x-1) -- (z-1);
\draw[dashed,->] (x-2) -- (z-1);
\draw[dashed,->] (x-2) -- (z-2);
\draw[dashed,->] (x-3) -- (z-2);
\draw[dashed,->] (x-3) -- (z-3);
\draw[dashed,->] (x-4) -- (z-3);
\draw[dashed,->] (x-4) -- (z-4);
\draw[dashed,->] (x-5) -- (z-4);
\draw[dashed,->] (x-5) -- (z-5);
\draw[dashed,->] (x-6) -- (z-5);
\draw[->] (x-6) -- node[left]{$\varepsilon_\ell$} (z-6);
\draw[->] (x-7) -- node[right]{$\varepsilon_r$} (z-6);
\draw[dashed,->] (x-7) -- (z-7);
\draw[dashed,->] (x-8) -- (z-7);
\node[below of=x-1, node distance=0.5cm] (w1) {\scriptsize\texttt{Double}};
\node[below of=x-2, node distance=0.5cm] (w2) {\scriptsize\texttt{isNaN}};
\node[below of=x-3, node distance=0.5cm] (w3) {\scriptsize\texttt{x}};
\node[below of=x-4, node distance=0.5cm] (w4) {\scriptsize\texttt{Double}};
\node[below of=x-5, node distance=0.5cm] (w5) {\scriptsize\texttt{isNaN}};
\node[below of=x-6, node distance=0.5cm] (w6) {\scriptsize\texttt{y}};
\node[below of=x-7, node distance=0.5cm] (w7) {\scriptsize\texttt{x}};
\node[below of=x-8, node distance=0.5cm] (w8) {\scriptsize\texttt{y}};

\node[output] (z-1-2) at (5,2) {$\hat{x}_5$};
\node[output] (z-2-2) at (6,2) {$\hat{x}_6$};
\draw[dashed,->] (z-6) -- node[left]{$\delta_\ell$} (z-1-2);
\draw[dashed,->] (z-6) -- node[right]{$\delta_r$} (z-2-2);
\end{tikzpicture}
\vspace{-0.25\baselineskip}
\caption{First iteration of encoding a stream of lexemes}
\vspace{-1.5\baselineskip}
\label{fig:ast.encode.greedy}
\end{figure}

The third and last stage of our machine learning phase involves clustering identifiers' embeddings.~%
For example, we clustered Math-63 identifiers' embeddings and used t-SNE~\cite{Maaten:2008} to reduce their dimensionality $\mathbb{R}^n\rightarrow\mathbb{R}^2$.~%
We plotted a handful of identifiers in our online appendix~\cite{onlineappendix} where terms with similar semantics appear to be proximate in the feature space.~%
In this context, by similar ``semantics,'' we mean terms are used in similar ways in the program (i.e., their token neighborhoods are similar).~%

Our machine learning phase induces models from source code.
These models are then used to make decisions.~%
During the program repair phase, decisions are made on whether or not to transform ingredients and \emph{we use the clusters to operationalize the decision criterion}.~%
When transforming ingredients, the only identifiers that may replace an out-of-scope variable access are identifiers in the same cluster (Sec.~\ref{sub:repair}).~%

\subsection{Program~Repair~Phase}
\label{sub:repair}

\subsubsection*{Core~repair~loop}

DeepRepair is based on a typical generate-and-validate repair loop à la GenProg.~%
First, it begins in a traditional way by running fault localization to get a list of suspicious statements and their suspicious values.~%
These suspicious statements serve as {modification points}, i.e., points where candidate patches would be applied.~%
For each modification point in sequence, a repair operator is used to transform that statement.~%
Then, one tries to recompile the changed class (since repair operators do not guarantee a well-formed program after modification), and the tentative patch is validated against the whole test suite.~%

\subsubsection*{Repair~operators}

In DeepRepair, the repair operators are ``addition of statement'' and ``replacement of statement.''~%
Contrary to GenProg and jGenProg, 
we do not use statement removal because it generates too many incorrect patches~\cite{Qi:2014}.~%
Addition and replacement are redundancy-based repair operators, since they need to select code from elsewhere in the codebase.~%
In DeepRepair, the reused code are in ``ingredient pools,'' with three main pools, corresponding to whether the ingredient is in the same class as the modification point (local reuse), in the same package (package reuse), or anywhere in the codebase (global reuse).~%
DeepRepair departs from jGenProg on two fundamental points.~%
First, while the default jGenProg operators randomly pick one statement from the ingredient pool, DeepRepair sorts the ingredients according to a specific criterion based on code similarity.~%
Second, the default jGenprog operators reuse code ``as is'' at the modification point, without applying any transformation, so it could happen that the ingredient has variables that are not in scope at the modification point (or wrongly typed), resulting in an \emph{uncompilable}---obviously incorrect---candidate patch.
\emph{On the contrary, DeepRepair has the ability to transform an ingredient so that it fits within the programming scope of the modification point.}~%
In DeepRepair, the sorting and transformation of ingredients is based on deep learning.~%
A combination of a sorting and transformation technique is called a \emph{fix space navigation strategy}.~%
In this paper, we explore five novel navigation strategies implemented in DeepRepair.~%

\subsubsection*{Sorting~ingredients}

When DeepRepair applies a repair operator, it sorts the available repair ingredients.~%
DeepRepair prioritizes the ingredients that come from methods (resp. classes) that are similar to the method (resp. class) containing the modification point.~
For example, if the suspicious statement is the return statement in Fig.~\ref{fig:method}, then DeepRepair takes the parent method MathUtils::equals(double,double) and uses the method-level similarity list to sort methods in the codebase.~%
Then, it extracts the statements 
from each similar method in order and enqueues them in a first-in first-out ingredient queue.~%

\subsubsection*{Transforming~ingredients}

For a patch to be compiled, the ingredient must ``fit'' in the modification point in the sense that all variable accesses must be in scope.~%
We refer to these ``fit'' ingredients as \emph{compilable} ingredients.~%
The key point of DeepRepair is its ability to transform ingredients, so that a repair ingredient can be adapted to a particular context, resulting in uncompilable ingredients becoming compilable at the modification point.~%
If a variable is out of scope, then DeepRepair examines the other identifiers \emph{in its cluster} to determine whether any of them are {in} scope.~%
For example, suppose a patch attempt consists of replacing the return statement in Fig.~\ref{fig:method} and suppose we poll the following
ingredient:~%
\texttt{\textcolor{blue}{return} equals(x,y,1) || FastMath.abs(y-x) <= eps;}
The variable {eps} is out of scope at the modification point, but SAFE\_MIN---a term in its cluster---{is} in scope, so we replace {eps} with {SAFE\_MIN} in the ingredient.~%
Applying this patch with a transformed ingredient yields a correct patch for Math-63 (Sec.~\ref{sub:rq4}).~%
Transforming repair ingredients using clusters based on learned embeddings enables DeepRepair to create novel patches. These patches would be impossible to generate if only raw ingredients were used as done in jGenProg.~
To emphasize the fact that the space of possible DeepRepair patches is \emph{not} the same as jGenProg, DeepRepair's fix space comprises \emph{every} statement in the codebase, which is jGenProg’s fix space.
However, the ability to transform ingredients means DeepRepair is able to generate patches that are not attainable by jGenProg.
DeepRepair's correct patch for Math-63 included a statement that did not exist in Math-63 (Revision ID: d2a5bc0); therefore, jGenProg is not capable of generating this correct patch.~%

\section{Empirical~Validation}
\label{sec:validation}

In Sec.~\ref{sub:experiment}, we define our research questions, define the goal of our empirical study, establish the experimental baseline configuration, and state our hypotheses.~%
We conclude our plan with our experimental design: comparative experiments are characterized by treatments, experimental units (i.e., the objects to which we apply the treatments), and the responses that are measured.~%
Then we describe our data collection procedures (Sec.~\ref{sub:data}) and conclude this section by specifying our analysis procedures for each one of our research questions (Sec.~\ref{sub:analysis}).~%

\subsection{Experiment~Scope~and~Plan}
\label{sub:experiment}

Our research questions included the following:~%
\begin{description}
\item[\emph{RQ1}]
Do code similarities based on deep learning improve fix space navigation as compared to a uniform random search strategy?~%
\item[\emph{RQ2}]
Does ingredient transformation using embeddings based on deep learning effectively transform repair ingredients as compared to 
a default ingredient application algorithm that does not transform ingredients?~%
\item[\emph{RQ3}]
Does DeepRepair, our learning-based approach that uses code similarities \emph{and} transforms 
ingredients, improve fix space navigation as compared to 
jGenProg?
\item[\emph{RQ4}]
Does DeepRepair generate higher quality patches than 
jGenProg?
\end{description}
The goal of our empirical study {was} to {analyze} 
ingredient search strategies for the {purpose} of evaluation with {respect} to 
effectiveness~\cite{Wohlin:2000}.~%
Our 
study was from the 
viewpoint of the software maintainer in the {context} of {six} open-source Java projects, a collection of (real) reproducible bugs, and a collection of JUnit test cases~\cite{Wohlin:2000}.~%

The baseline configuration, jGenProg, was the uniform random search strategy where ingredients were selected from a pool of equiprobable statements.~%
We configured the baseline with a cache so the same modification instance (i.e., ingredient and operator instance) was never {attempted} more than once to improve its efficiency.~%
The baseline was also configured with a default ingredient application algorithm that analyzed the variable accesses in a repair ingredient, matching accesses' names and types to variables in scope.~%
If at least one variable access failed to match a variable in scope, then the ingredient was discarded.~%
Tab.~\ref{tab:configurations} lists the DeepRepair configurations.~
%
%
%
%
Each configuration evaluates sorting at a particular level of granularity (executable- or type-level) with or without the ability to transform ingredients.~%

We formalized our experiment into the following hypotheses:~%
\begin{description}
\item[$H_0^{1a}$]
Using code similarities generates the same number of test-adequate patches (on average) as jGenProg.~%
\item[$H_0^{1b}$]
Using code similarities attempts\footnote{An \textbf{attempt} is defined to be a request sent to the fix space for an ingredient.} 
the same number of ingredients (on average) before finding a {test-adequate patch} as jGenProg.~%
\item[$H_0^{2a}$]
Ingredient transformation using embeddings generates the same number of test-adequate patches (on average)
as jGenProg.~%
\item[$H_0^{2b}$]
Ingredient transformation using embeddings attempts the same number of ingredients (on average)
before finding a {test-adequate patch} as jGenProg.~%
\item[$H_0^{3a}$]
DeepRepair generates the same number of {test-adequate patches} (on average) as jGenProg.~%
\item[$H_0^{3b}$]
DeepRepair attempts the same number of ingredients (on average) before finding a {test-adequate patch} as jGenProg.~%
\item[$H_0^4$]
{There is no significant difference in quality 
between patches generated by DeepRepair and jGenProg.}~%
\end{description}
%
%

\begin{table}[t]
\caption{DeepRepair~configurations~evaluated~in~the~study}
\vspace{-0.5\baselineskip}
\label{tab:configurations}
\centering
\begin{tabular}{@{}ll@{}}
\toprule
Code					& Features\\
\midrule
\multirow{2}{*}{ED}		& (E)xecutable-level similarity ingredient sorting\\
						& (D)efault ingredient application (no ingredient transformation)\\
\addlinespace
\multirow{2}{*}{TD}		& (T)ype-level similarity ingredient sorting\\
						& (D)efault ingredient application (no ingredient transformation)\\
\addlinespace
\multirow{2}{*}{RE}		& (R)andom ingredient sorting\\
						& (E)mbeddings-based ingredient transformation\\
\addlinespace
\multirow{2}{*}{EE}		& (E)xecutable-level similarity ingredient sorting\\
						& (E)mbeddings-based ingredient transformation\\
\addlinespace
\multirow{2}{*}{TE}		& (T)ype-level similarity ingredient sorting\\
						& (E)mbeddings-based ingredient transformation\\
\bottomrule
\end{tabular}
\vspace{-1.5\baselineskip}
\end{table}
We chose the following {dependent variables}:~number of test-adequate patches and
number of ingredients attempted.~
We chose the following {independent variables} (emphasizing the factors):~
\emph{ingredient~search~strategy}, \emph{scope}, \emph{clone granularity}, \emph{variable resolution algorithm}, fault localization threshold, maximum number of suspicious candidates, and programming language.~%
The fault localization threshold was fixed at 0.1;~%
%
%
%
%
the maximum number of suspicious candidates was fixed at 1,000;~%
%
%
%
%
and the language was fixed on Java.~%
Fixing the language on Java precluded comparing our approach to other popular repair approaches such as Prophet~\cite{Long:2016}, Angelix~\cite{Mechtaev:2016}, and SearchRepair~\cite{Ke:2015} that target C programs.~%
The random seed for each experimental configuration ranged from one to {three}.~%


%
%
We designed comparative experiments to measure the statistical significance of our results.~%
Typically, empirical studies in the field report the number of test-adequate patches found, but we conducted a large-scale study, report the number of test-adequate patches found, and measure the statistical significance of differences to contextualize the results.~%
{For our quantitative study,} the treatments in our experimental design were the ingredient search strategies.~%
%
%
The experimental units were the buggy program revisions, and the responses were the number of test-adequate patches found and the number of ingredients attempted before finding a {test-adequate patch}.~%
%
%


\subsection{Data~Collection~Procedure}
\label{sub:data}

The first stage of our recognition phase involved creating a model of source code (Sec.~\ref{sub:recognition}).~%
We used Spoon~\cite{Pawlak:2015}, an open-source library for analyzing and transforming Java source code, to build a model for each buggy program revision.~%

Given the 
models, we implemented program processors for querying program elements at three levels of granularity:~file-, type-, and executable-level granularity.~%
Types included classes and interfaces, and executables included methods and constructors, but we omitted anonymous blocks.~%
We only queried top-level types and executables to control the number of similarities to be computed.~%
In this context, by ``top-level,'' we mean types or executables that did not have a parent type or executable, respectively.~%
For example, a nested class would not be included, but its enclosing class may be included.~%
Then we extracted the yield~\cite{Aho:2006} from each program element's {syntax} tree along with 
a key for uniquely identifying the element to build three corpora (Sec.~\ref{sub:recognition}).~%
So each line of a type-level corpus corresponded to a class or interface in the program.~%
The only normalization we performed was replacing characters, floats, integers, and strings with their respective type.~%

The first stage of our learning phase involved inducing neural network language models from the normalized file-level corpora (Sec.~\ref{sub:learning}).~%
We used word2vec~\cite{Mikolov:2013:ICLR,Mikolov:2013:NIPS,Mikolov:2013} to learn embeddings for \emph{each} buggy program revision.~
We selected word2vec over other architectures because the models can be trained quickly, and we only used the language models' embeddings to initialize the embeddings for the recursive autoencoders rather than randomly initialize them.~%
We used the skip-gram model and set the size of the word vectors to 400 in accordance with previous studies using similar subject systems~\cite{White:2016}.~%
We set the maximum skip length between words to 10, used a hierarchical softmax to optimize the computation of output vectors' updates~\cite{Rong:2014}, and trained each model for 20 iterations.~%
The language models enabled us to transform the file-level corpus for each program revision into streams of embeddings.~%

Then we trained recursive autoencoders to encode streams of embeddings.~%
The encoders used hyperbolic tangent activations (i.e., $f\defeqq\text{tanh}$ in Eq.~\eqref{eq:z}), used L-BFGS~\cite{Nocedal:1980} to optimize costs in batch mode, and trained for up to 50 epochs.~%
After an encoder was trained on a revision's file-level corpus, we used it to encode every type and executable in the revision's type- and executable-level corpora, respectively.~%
Given the encodings, we computed the pairwise Euclidean distance between each pair of types in the type-level corpus and each pair of executables in the executable-level corpus to measure similarities for each program revision.~%

Next, we extracted the term embeddings from the trained encoder and clustered them using $k$-means.~%
For each revision, to determine $k$, we used simulated annealing, initializing both $k$ and the temperature to be the square root of the corresponding vocabulary size.~%
The reason we chose the square root of the vocabulary size is because of its effectiveness in related contexts that categorize words~\cite{Mikolov:2011:RNNLM}.~%
The objective we optimized was minimizing the number of points with negative silhouette values.~%
Thus, at the end of the learning phase, each buggy revision had a cached list of executable- and type-level similarities as well as a categorization for identifiers.~%

The final phase of our technical approach involved automatically repairing buggy program revisions (Sec.~\ref{sub:repair}).~%
Our subject systems comprised {six} open-source Java projects including {374} buggy program revisions in Defects4J database version {1.1.0}~\cite{Just:2014,Just:2014:ISSTA}.\footnote{We could not build the Spoon model for Mockito bugs 1--21, which was likely because of missing or incompatible dependencies.}~%
In Tab.~\ref{tab:projects}, we report median values since each project has several buggy program revisions.~%

To run 
repair experiments, we used Astor~\cite{Martinez:2016:ASTOR}, an automatic software repair framework for Java.~%
{Within the Astor framework, we leveraged GZoltar~\cite{Campos:2012}, a spectrum-based fault localization tool, to compute the Ochiai formula~\cite{Abreu:2006} for statements' suspicious values.}~%

Each trial corresponded to a seeded treatment, which was a factorial of strategy and scope.~%
Empirical studies indicated that fragment locality matters in software maintenance and evolution, so we analyzed three different levels of scope:~local, package, and global~\cite{Martinez:2014}.~%
{For local scope, Astor builds the ingredient search space by amalgamating the distinct set of classes that contain at least one suspicious statement.}~%
{For package scope, Astor computes the distinct set of packages that contain at least one suspicious statement and builds the ingredient search space using the set of classes in those packages.}~%
{For global scope, Astor builds the ingredient space using all classes from the application under repair.}~%
{We configured Astor to \emph{not} stop at the first patch found and---starting from the original program---to continue searching for other patches until reaching a three-hour time limit.}~%
In total, we ran {20,196} trials ({374} buggy program revisions $\times$ 6 search strategies $\times$ 3 levels of scope $\times$ {3} random seeds) on subclusters comprising 64 compute nodes running Red Hat Enterprise Linux~6.2.~
Each compute node was a Dell PowerEdge C6100 serving two Intel Xeon~X5672 quad-core processors at 3.2~GHz with 12~MB L3~cache and at least 48~GB of 1333~MHz main memory.~%
Each trial was allocated {two} (hyper-threaded) cores and 
evolved one program variant for three hours using three repair operators:~InsertAfterOp, InsertBeforeOp, and ReplaceOp.~%
\emph{We did not include RemoveOp in our operator space because we only focused on repair operators that reuse code}.~%
We also wanted to guard against meaningless patches that simply remove functionality.~%

\begin{table}[t]
\caption{Project~statistics}
\vspace{-0.5\baselineskip}
\label{tab:projects}
\centering
\begin{tabular}{@{}lcr@{,}lr@{,}lc@{}}
\toprule
Project				& Files	& \multicolumn{2}{c}{LOC} & \multicolumn{2}{c}{Tokens} & Vocab.\\
\midrule
Apache~commons-lang	& 221	& 48&890	& 420&000	& \phantom{2}4,672\\
Apache~commons-math & 845	& 97&130	& 830&000	& \phantom{2}8,450\\
Closure~compiler 	& 937	& 247&300	& 1,449&000	& 26,490\\
JFreechart 			& 952	& 130&300	& 921&800	& \phantom{2}9,008\\
Joda-Time			& 316	& 81&640	& 736&300	& \phantom{2}5,989\\
Mockito				& 680	& 44&990	& 309&500	& \phantom{2}5,735\\
\bottomrule
\end{tabular}
\vspace{-1.5\baselineskip}
\end{table}
\subsection{Analysis~Procedure}
\label{sub:analysis}

\subsubsection*{RQ1}

We analyzed the effectiveness of fix space navigation strategies in two parts.~%
The {first} part analyzed 
effectiveness at generating test-adequate patches.~%
We used the non-parametric Wilcoxon test with a Bonferroni correction to compare the number of test-adequate patches using 
jGenProg versus 
the strategy using code similarities at executable- and type-level granularity (i.e., ED and TD in Tab.~\ref{tab:configurations}).~%
We used Wilcoxon since the test-adequate patch counts could be paired, and we used a Bonferroni correction since we performed several tests simultaneously at different levels of scope and strategy.~%
{To complement our statistical analysis on number of test-adequate patches found, we also computed the difference between the set of jGenProg patches and the set of DeepRepair patches.}~%
Specifically, if $D$ is the set of DeepRepair patches, and $J$ is the set of jGenProg patches, then we computed 
$\left\vert{D\setminus J}\right\vert/\left\vert{D}\right\vert$, the percentage of DeepRepair patches that were not found by jGenProg.~
The {second} part analyzed the number of attempts to generate test-adequate patches.~%
We used the non-parametric Mann-Whitney test with a Bonferroni correction to compare the number of attempts to generate test-adequate patches using jGenProg and 
using code similarities at executable- and type-level granularity.~%
To complement our analysis, we also plotted the number of attempts to generate {compilable} ingredients.~

\subsubsection*{RQ2}

We analyzed effectiveness in two parts.~%
We used the Wilcoxon test to compare the number of test-adequate patches using jGenProg versus the uniform random search strategy with the embeddings-based 
ingredient transformation algorithm (i.e., RE in Tab.~\ref{tab:configurations}).~%
{The algorithm gives jGenProg the ability to transform repair ingredients containing variable accesses that are out of scope.}~%
{We also computed the difference between the set of jGenProg patches and the set of DeepRepair patches.}~%
Additionally, we used the Mann-Whitney test to compare the number of attempts to generate test-adequate patches and plotted the number of attempts to generate compilable ingredients.~%

\subsubsection*{RQ3}

Our experimental design for \emph{RQ3} was virtually identical to our design for \emph{RQ1} except here we compared jGenProg to the strategy using code similarities with the embeddings-based 
ingredient transformation algorithm (i.e., EE and TE in Tab.~\ref{tab:configurations}).~%

\subsubsection*{RQ4}

We used correctness as a proxy for quality.~%
Three judges evaluated the same random sample of 30 (15 jGenProg and 15 DeepRepair) patches to assess correctness.~%
Martinez et al.~\cite{Martinez:2015} defined the \emph{correctness} of a patch to be one of three values:~correct, incorrect, or unknown.~%
\emph{Correct} denotes a patch is equivalent (according to the judge's understanding) to the human-written patch.~%
Judges were also prompted for their confidence in their correctness rating where confidence was one of four values:~high, moderate, slight, and none.~%
Additionally, for reproducibility, judges also assessed the readability of each patch, where the \emph{readability} of a patch was either easy, medium, or hard in accordance with previous studies on patch correctness~\cite{Qi:2015,Martinez:2016}.~%
We define readability to be the subjective qualification of how easily the patch can be understood.~%
Readability is a subjective aggregation of different quantitative metrics: patch size in number of lines, number of involved variables, number of method calls, and the types of AST elements being inserted or replaced.~%
There is no accepted quantitative aggregation of these metrics, and we think that a quantitative aggregation would be project- and even bug-dependent.~%
In addition to the random sample, judges also evaluated the patches generated by jGenProg that were not found by DeepRepair and patches generated by DeepRepair that were not found by jGenProg.~%
\section{Empirical~Results}
\label{sec:results}

\begin{table}[t]
\caption{Patches~found~at~(L)ocal,~(P)ackage,~and~(G)lobal~scope}
\vspace{-0.5\baselineskip}
\label{tab:patches}
\centering
\begin{tabular}{@{}lccccccc@{}}
\toprule
ProjectID					& BugID	& jGenProg & ED & TD & RE & EE & TE\\
\midrule
\multirow{12}{*}{Chart} 	& \phantom{0}1	& LPG	& LPG	& LPG	& LPG	& LPG	& LPG\\
							& \phantom{0}3	& LPG	& LPG	& LPG	& LPG	& LPG	& LPG\\
							& \phantom{0}5	& LPG	& LPG	& LPG	& LPG	& LPG	& LPG\\
							& \phantom{0}7	& LPG	& LPG	& LPG	& LPG	& LPG	& LPG\\
							& \phantom{0}9	& -	& -	& -	& -	& \phantom{L}P\phantom{G}	& -\\
							& 12	& \phantom{LP}G	& LPG	& \seedtwo{LPG}	& \phantom{LP}\seedtwo{G}	& LPG	& \seedtwo{LPG}\\
							& 13	& LPG	& LPG	& LPG	& LPG	& LPG	& LPG\\
							& 14	& LPG	& LPG	& LPG	& LPG	& LPG	& LPG\\
							& 15	& LPG	& LPG	& LPG	& LPG	& LPG	& LPG\\
							& 18	& \seedtwo{L}\phantom{PG}	& -	& -	& L\phantom{PG}	& -	& -\\
							& 25	& LPG	& L\seedtwo{PG}	& \seedthr{LPG}	& LPG	& LPG	& -\\
							& 26	& LPG	& LPG	& LPG	& LPG	& LPG	& \seedtwo{L}PG\\
\midrule
\multirow{8}{*}{Lang}		& \phantom{0}7	& LPG	& LPG	& LPG	& LPG	& LPG	& LPG\\
							& 10	& LPG	& L\seedtwo{P}G	& L\seedtwo{P}G	& L\seedthr{P}\seedtwo{G}	& LPG	& LP\seedtwo{G}\\
							& 20	& LPG	& LPG	& LPG	& LPG	& LPG	& LPG\\
							& 22	& LPG	& LPG	& LPG	& LPG	& LPG	& LP\phantom{G}\\
							& 24	& LPG	& LPG	& L\phantom{PG}	& LPG	& L\phantom{PG}	& L\phantom{PG}\\
							& 27	& LPG	& LPG	& LPG	& LPG	& LPG	& LPG\\
							& 38	& \phantom{L}PG	& -	& -	& \phantom{L}PG	& -	& -\\
							& 39	& LPG	& LPG	& LPG	& \seedtwo{LPG}	& \seedtwo{L}\phantom{PG}	& -\\
\midrule
\multirow{31}{*}{Math}		& \phantom{0}2	& LPG	& LPG	& LPG	& LPG	& LPG	& LPG\\
							& \phantom{0}5	& LPG	& LPG	& LPG	& LPG	& LPG	& LPG\\
                            & \phantom{0}6	& L\seedthr{P}\phantom{G}	& LPG	& LPG	& LP\phantom{G}	& \seedthr{LP}\phantom{G}	& \phantom{LP}\seedthr{G}\\
                            & \phantom{0}\seedtwo{7}	& \seedtwo{L}\phantom{PG}	& -	& -	& -	& -	& -\\
                            & \phantom{0}8	& -	& -	& -	& \phantom{LP}G	& LPG	& LPG\\
							& 18	& \seedtwo{L}\phantom{PG}	& -	& -	& \phantom{L}P\phantom{G}	& -	& -\\
                            & 20	& LP\seedtwo{G}	& LPG	& LPG	& LPG	& LP\seedtwo{G}	& LP\seedtwo{G}\\
                            & 22	& L\seedtwo{PG}	& LPG	& -	& L\seedtwo{P}G	& LPG	& -\\
                            & 24	& -	& -	& -	& LP\phantom{G}	& -	& -\\
                            & 28	& LPG	& LPG	& LPG	& LPG	& LPG	& LPG\\
                            & 32	& \seedthr{L}\phantom{PG}	& \seedthr{LPG}	& -	& \seedtwo{L}P\phantom{G}	& -	& -\\
                            & 40	& LPG	& L\seedthr{P}G	& LP\phantom{G}	& LPG	& LPG	& \seedthr{L}\phantom{PG}\\
                            & \seedtwo{44}	& \phantom{L}\seedtwo{P}\phantom{G}	& -	& -	& -	& -	& -\\
                            & 49	& LPG	& LPG	& LPG	& LPG	& LP\seedtwo{G}	& LPG\\
                            & 50	& LPG	& LPG	& LP\phantom{G}	& LPG	& LP\phantom{G}	& LP\phantom{G}\\
                            & 53	& LPG	& LPG	& LPG	& LPG	& LPG	& LPG\\
                            & 56	& L\phantom{PG}	& LP\seedtwo{G}	& \seedtwo{L}\seedthr{P}\phantom{G}	& L\seedtwo{PG}	& \seedtwo{L}P\phantom{G}	& -\\
                            & 57	& \phantom{LP}G	& L\phantom{PG}	& LP\phantom{G}	& \phantom{LP}G	& -	& LP\phantom{G}\\
                            & \seedtwo{58}	& -	& -	& -	& \seedtwo{L}\seedthr{P}\phantom{G}	& -	& -\\
                            & 60	& \phantom{LP}G	& LP\phantom{G}	& LPG	& \phantom{LP}G	& LP\phantom{G}	& LP\phantom{G}\\
                            & 63	& -	& -	& -	& LPG	& LPG	& LPG\\
                            & 64	& L\phantom{PG}	& -	& -	& -	& -	& -\\
                            & 70	& LPG	& LPG	& LPG	& LPG	& LPG	& LPG\\
                            & 71	& L\seedtwo{P}G	& -	& -	& \seedtwo{LP}G	& -	& -\\
                            & 73	& LPG	& LPG	& LPG	& LPG	& LPG	& LPG\\
                            & 74	& \phantom{L}PG	& -	& -	& \phantom{L}P\seedtwo{G}	& -	& -\\
                            & 77	& LPG	& \seedtwo{L}\phantom{PG}	& LPG	& LPG	& L\phantom{PG}	& LPG\\
                            & 78	& LPG	& LPG	& LPG	& LPG	& LPG	& LPG\\
                            & 80	& LPG	& LPG	& LPG	& LPG	& \phantom{L}\seedthr{P}\phantom{G}	& \seedthr{L}\seedtwo{PG}\\
                            & 81	& LPG	& LPG	& LPG	& LPG	& LPG	& LPG\\
                            & 82	& -	& -	& -	& -	& \phantom{L}PG	& \phantom{LP}G\\
                            & 84	& LPG	& LP\phantom{G}	& LP\phantom{G}	& LPG	& LP\phantom{G}	& LP\phantom{G}\\
                            & 85	& LPG	& LPG	& LPG	& LPG	& LPG	& LPG\\
                            & 98	& LPG	& LPG	& LPG	& LPG	& LP\seedtwo{G}	& LPG\\
\addlinespace
Total						& 54	& 48	& 40	& 38	& 49	& 42	& 38\\
\bottomrule
\end{tabular}
\vspace{-1.5\baselineskip}
\end{table}

%
%
%
%
%
%
%
%
%
%
\dissertation{DEFN Main effect. The effect of one factor, averaged across the levels of any other independent factor, on a response.}
\dissertation{DEFN Factor interactions. occur when the effect of a factor depends on the level setting of another factor.}
We ran {20,196} repair trials, {247} of which were killed.~%
The {19,949} trials that finished took a total of {2,616} days of computation time.~%
{Zero patches were found for Closure, Mockito, and Time bugs.}~%
The baseline configuration found test-adequate patches for {48} different bugs.~%
{Six} other bugs were unlocked by DeepRepair configurations.
The trials found {19,832} different test-adequate patch \emph{instances} and attempted {406,443,249} ingredients.~%

\begin{figure*}
\centering
\includegraphics[width=0.99\textwidth]{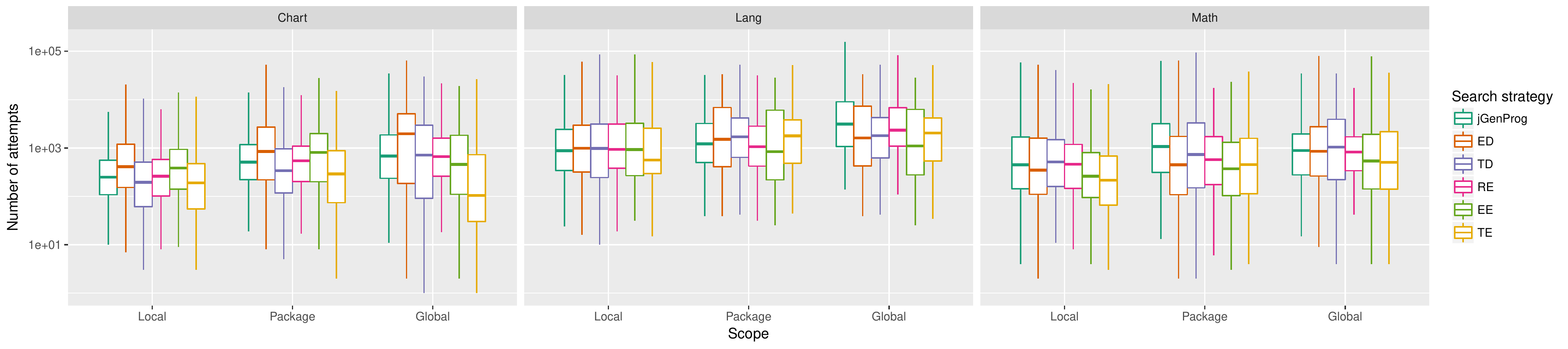}
\vspace{-0.5\baselineskip}
\caption{Number of attempts to find a {test-adequate} patch}
\vspace{-1.5\baselineskip}
\label{fig:patch_attempts}
\end{figure*}

\subsection{RQ1 (Analysis of ED and TD Strategies)}

Tab.~\ref{tab:patches} lists the bugs for which test-adequate patches were found at (L)ocal, (P)ackage, and (G)lobal scope.~%
jGenProg found test-adequate patches for {48} bugs.~%
The treatments ED and TD found patches for {40} and {38} bugs, respectively.~%

Since many configurations found more than one patch for the same bug, we compared the patch counts between jGenProg and the two treatments ED and TD to see whether one configuration was more productive than another.~%
We failed to reject the null hypothesis {$H_0^{1a}$} at each level of scope.~%
Next, we analyzed the sets of patches and observed approximately {99\%, 25\%, and 36\% of DeepRepair's patches for Chart, Lang, and Math} were not found by jGenProg.~%
This result means that DeepRepair finds alternative patches.~%

We also counted the number of attempts to find each patch.
Fig.~\ref{fig:patch_attempts} shows descriptive statistics for the number of attempts to find a test-adequate patch.~%
We failed to reject the null hypothesis $H_0^{1b}$ at each level of scope.~%
We also counted the number of attempts to find each compilable ingredient (as defined in Sec.~\ref{sub:repair}).~%
Fig.~\ref{fig:ing_attempts} shows descriptive statistics for the number of attempts to find a compilable ingredient at each level of scope for jGenProg, ED, and TD.~%
Generally, sorting the fix space using code similarities results in fewer attempts before finding a compilable ingredient.~%
\\\textbf{Key~result.}~DeepRepair's search strategy using code similarities generally finds compilable ingredients faster than the baseline, but this improvement neither yields test-adequate patches in fewer attempts (on average) nor finds significantly more patches (on average) than the baseline.~%
{Yet there were notable differences between DeepRepair and jGenProg patches.}~%

\subsection{RQ2 (Analysis of RE Strategy)}

The treatment RE found test-adequate patches for {49} bugs.
%
%
Comparing the patch counts, we failed to reject the null hypothesis {$H_0^{2a}$} at each level of scope.~%
Analyzing the set difference, {53\%, 3\%, and 53\% of DeepRepair's patches for Chart, Lang, and Math} were not found by jGenProg.~%
We failed to reject the null hypothesis {$H_0^{2b}$} at each level of scope.~%
This can also be graphically seen in Fig.~\ref{fig:patch_attempts} that shows descriptive statistics for the number of attempts to find a test-adequate patch for jGenProg and RE.~%
\\\textbf{Key~result.}~DeepRepair's search strategy using the embeddings-based ingredient transformation algorithm neither yields 
patches in fewer attempts (on average) nor finds significantly more patches (on average) than jGenProg, but there were notable differences between DeepRepair and jGenProg patches.~%

\subsection{RQ3 (Analysis of EE and TE Strategies)}

The treatments EE and TE found test-adequate patches for {42} and {38} bugs, respectively.~%
Comparing the patch counts, we failed to reject the null hypothesis {$H_0^{3a}$} at each level of scope.~%
Analyzing the set difference, 99\%, 28\%, and 51\% of DeepRepair's patches for Chart, Lang, and Math were not found by jGenProg.~%
Fig.~\ref{fig:patch_attempts} shows descriptive statistics for the number of attempts to find a test-adequate patch for jGenProg, EE, and TE.~%
We failed to reject the null hypothesis {$H_0^{3b}$} at each level of scope.~%
Fig.~\ref{fig:ing_attempts} shows descriptive statistics for the number of attempts to find a compilable ingredient at each level of scope for jGenProg, EE, and TE.~%
\\\textbf{Key~result.}~DeepRepair's search strategy using code similarities with the embeddings-based ingredient transformation algorithm generally finds compilable ingredients faster than jGenProg, but this improvement neither yields test-adequate patches in fewer attempts (on average) nor finds significantly more patches (on average) than jGenProg.~%
Once more, DeepRepair appears to find a complementary set of patches.~%

\subsection{RQ4 (Manual Assessment)}
\label{sub:rq4}

\begin{figure*}
\centering
\includegraphics[width=0.99\textwidth]{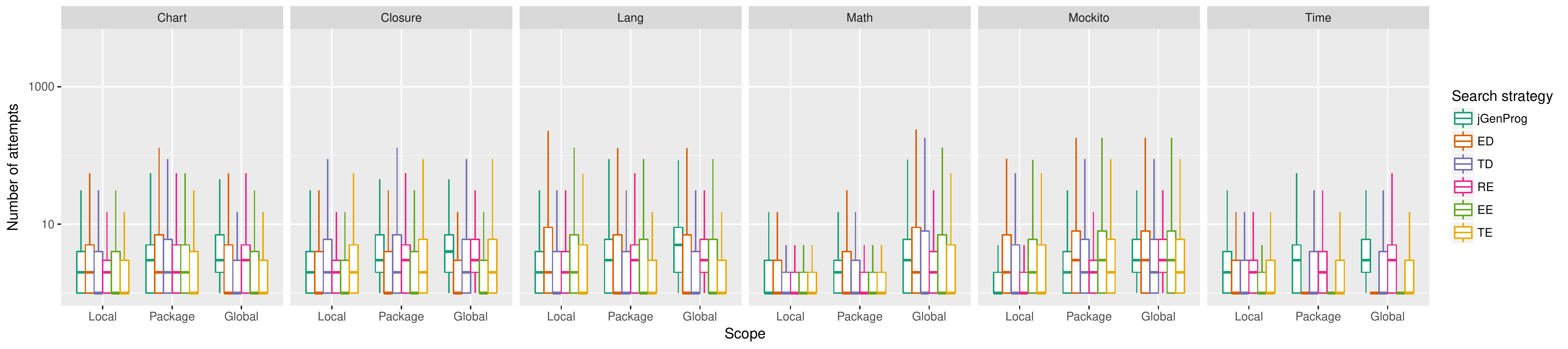}
\vspace{-0.5\baselineskip}
\caption{Number of attempts to find a compilable ingredient}
\vspace{-1.5\baselineskip}
\label{fig:ing_attempts}
\end{figure*}

After independently evaluating each sample patch, three judges discussed and resolved conflicts in terms of correctness.
Five DeepRepair patches and five jGenProg patches were evaluated to be correct.~%
We failed to reject the null hypothesis $H_0^4$.~%
Although judges did notice differences in the patches generated by the approaches, no significant difference in readability was reported.~%

In addition to the random sample, we manually examined the following specific sets of patches.~%
There were {three} bugs patched exclusively by jGenProg.~
We examined jGenProg's patches for these bugs and found that all of them were {clearly} incorrect.~%
On the other hand, there were {six} bugs patched exclusively by DeepRepair configurations.~
We also examined these patches and report some of our findings below.~%

\noindent\textbf{Chart-9.}~The human-written patch was~%
\begin{lstlisting}[basicstyle=\ttfamily\scriptsize]
+ if (endIndex < 0 || endIndex < startIndex) {
- if (endIndex < 0) {
\end{lstlisting}
None of the {identifiers} in the human-written patch were new (cf. Sec.~\ref{sub:redundancy}), but the conditional {expression} 
\begin{lstlisting}[basicstyle=\ttfamily\scriptsize]
  (endIndex < 0 || endIndex < startIndex)
\end{lstlisting}
{was} novel w.r.t.~the codebase, so generate-and-validate techniques that cannot generate new code would never find this patch.~%
The selection statement that DeepRepair generated passed the test suite, but it was incorrect (to the best of our knowledge).~%
Notably, the DeepRepair patch contained the conditional expression 
\begin{lstlisting}[basicstyle=\ttfamily\scriptsize]
  (endIndex < 1 || endIndex > LAST_WEEK_IN_YEAR)
\end{lstlisting}
whose syntactic structure resembled the human-written expression.~%
DeepRepair's conditional expression was novel w.r.t.~the codebase as it was generated by transforming an ingredient.~%
The expression was originally 
\begin{lstlisting}[basicstyle=\ttfamily\scriptsize]
  (result < 1 || result > LAST_WEEK_IN_YEAR)
\end{lstlisting}
but DeepRepair recognized \emph{similarities} in how the identifiers, {result} and {endIndex}, were used in the codebase, so it replaced result---a variable out of scope---with {endIndex}.~%

\noindent\textbf{Math-63.}~The human-written patch was~%
\begin{lstlisting}[basicstyle=\ttfamily\scriptsize]
  public static boolean equals(double x, double y)
-    return Double.isNaN(x) && Double.isNaN(y) || x == y;
+    return equals(x,y,1);
\end{lstlisting}
Again, none of the identifiers in the human-written patch were new, but the statement was novel w.r.t.~the codebase.~%
DeepRepair generated the following patch:~%
\begin{lstlisting}[basicstyle=\ttfamily\scriptsize]
  public static boolean equals(double x, double y)
-    return Double.isNaN(x) && Double.isNaN(y) || x == y;
+    return equals(x,y,1) || FastMath.abs(y-x) <= SAFE_MIN;
\end{lstlisting}
The ingredient was selected from a \emph{similar} method:~%
\begin{lstlisting}[basicstyle=\ttfamily\scriptsize,breaklines=false]
public static boolean equals(double x, double y, double eps)
   return equals(x,y,1) || FastMath.abs(y-x) <= eps;
\end{lstlisting}
However, the variable \lstinline[basicstyle=\normalsize]{eps} in the ingredient was not in scope at the modification point.~%
DeepRepair recognized similarities between how the identifiers \lstinline[basicstyle=\normalsize]{eps} and \lstinline[basicstyle=\normalsize]{SAFE_MIN} are used in the codebase, so it replaced \lstinline[basicstyle=\normalsize]{eps} with \lstinline[basicstyle=\normalsize]{SAFE_MIN}.~%
As a result, both the human-written patch and DeepRepair's patch invoke \lstinline[basicstyle=\normalsize]{equals(x,y,1)} which returns true if \lstinline[basicstyle=\normalsize]{x} and \lstinline[basicstyle=\normalsize]{y} are equal or within the range of allowed errors (inclusive).~%
In this case, the range of allowed error is defined to be zero floating point numbers between the two values, so the values must be the same floating point number or adjacent floating point numbers.~%
Therefore, when \lstinline[basicstyle=\normalsize]{equals(x,y,1)} is true, both the human-written patch and the DeepRepair patch return true since the conditional expression in the DeepRepair patch short-circuits.~%
When \lstinline[basicstyle=\normalsize]{equals(x,y,1)} is false, the human-written patch returns false, and since \lstinline[basicstyle=\normalsize]{SAFE_MIN} is defined to be the smallest normalized number in IEEE 754 arithmetic (i.e., 0x1.0p-1022), the DeepRepair patch returns false.~%
Hence, the DeepRepair patch is semantically equivalent to the human patch and considered correct.~%
\\\textbf{Key~result.}~There are apparent, critical differences observed in DeepRepair's patches compared to jGenProg, which unlock new bugs---that would otherwise have not been patched---by reusing \emph{and transforming} repair ingredients.~%
Our future work aims to extensively analyze more results to understand which defect \emph{classes} can be unlocked with DeepRepair's search strategies.~%
\section{Threats~to~Validity}
\label{sec:threats}



\subsubsection*{Internal~validity}

DeepRepair relies on deep learning to compute similarities among code elements at different levels of granularity.~%
Learning-based code clone detection has been evaluated at multiple levels of granularity with promising results~\cite{White:2016}.~%
Additionally, we manually examined small random samples of similar code fragments at executable- and type-level granularity from each project to validate some degree of textual/functional similarity in accordance with previous studies.~%
However, we do not claim to have used optimal settings for training on each program revision or even each project.
We also acknowledge the confounding configuration choice problem~\cite{Wang:2013}.~%
We did not adopt arbitrary configurations and tried to {justify} each configuration in our approach.~%

\subsubsection*{External~validity}

In our experiments we evaluate DeepRepair on 374 buggy program revisions (in six unique software systems) from the Defects4J benchmark.~
One threat is that the number of bugs may not be large enough to represent the actual differences between DeepRepair and jGenProg.~%

\subsubsection*{Construct~validity}

Our empirical evaluation is similar to all previous studies on program repair in that 
DeepRepair and jGenProg do not target buggy program revisions with multiple faults.~
Also, prior work noted some impact of flaky test cases on program repair~\cite{Martinez:2016}.~%
While we did not detect any flaky tests in our benchmarks during the experiments, their potential presence could impact both DeepRepair and jGenProg.~%

\subsubsection*{Conclusion~validity}

DeepRepair and jGenProg have random components, so different runs would possibly produce different patches.
However, our experiments were as computationally extensive as they could have 
been within our means, consisting of 19,949 trials spanning 2,616 days of computation time.~%
We could not manually analyze all the generated patches as this would require years of manual work, but we randomly sampled a subset for manual evaluation.~%
Three judges inspected 
each sample to minimize bias.~%
\section{Conclusion}
\label{sec:conclusion}

We introduced a novel learning-based algorithm to intelligently select and {transform} repair ingredients in a generate-and-validate repair loop based on the redundancy assumption. 
DeepRepair takes a novel perspective on the ingredient selection problem: it selects ingredients from similar methods or classes 
where similarity has been inferred with deep unsupervised learning.~%
Many repairs need 
to replace identifiers 
to make an ingredient compilable at a specific modification point.
To the best of our knowledge, DeepRepair is the first approach that expands the fix space by transforming ingredients using identifiers' similarities.~%
We conducted a computationally intensive empirical study and found that DeepRepair did not significantly improve effectiveness using a new metric (number of attempts), but DeepRepair did generate many patches that cannot be generated by existing redundancy-based repair techniques.~%
\section*{Acknowledgment}
We thank David Nader Palacio {for his help with conducting the empirical study}.~%
This work was performed 
using computing facilities at the College of William and Mary which were provided by contributions from the National Science Foundation, the Commonwealth of Virginia Equipment Trust Fund, and the Office of Naval Research.~%
This material is based upon work supported by the National Science Foundation under Grant No.~1525902.~%
This work was partially supported by the Wallenberg Artificial Intelligence, Autonomous Systems and Software Program (WASP) funded by Knut and Alice Wallenberg Foundation.
\bibliographystyle{unsrt}
\bibliography{thesis2}
\end{document}